\documentclass[%
  reprint,
  showpacs,preprintnumbers,
  amsmath,amssymb,
  aps,
  prb,
uplatex
]{revtex4-2}

\usepackage{graphicx}[]
\usepackage{dcolumn}
\usepackage{bm}
\usepackage{multirow}
\usepackage{braket}
\usepackage{color}
\usepackage{ulem}

\usepackage{hyperref}
\usepackage{url}

\begin{document}

\preprint{APS/123-QED}

\title{
Skyrmion crystal under $D_{3h}$ point group:\\ Role of out-of-plane Dzyaloshinskii-Moriya interaction
}

\author{Satoru Hayami$^{1,2}$ and Ryota Yambe$^{1}$}
\affiliation{
$^{1}$Department of Applied Physics, The University of Tokyo, Tokyo 113-8656, Japan\\
$^{2}$Faculty of Science, Hokkaido University, Sapporo 060-0810, Japan
}
 
\begin{abstract}
The Dzyaloshinkii-Moriya (DM) interaction that originates from relativistic spin-orbit coupling in noncentrosymmetric magnets is a source of topological spin textures. 
We theoretically investigate the possibility of a skyrmion crystal by focusing on the role of the out-of-plane DM interaction that is different from the polar- and chiral-type DM interaction. 
By performing the simulated annealing for a spin model on a triangular lattice belonging to the $D_{3h}$ point group, we construct low-temperature magnetic phase diagrams in an applied magnetic field. 
As a result, we find the instability toward different skyrmion crystals under in-plane and out-of-plane magnetic fields, where the key ingredients to stabilize the SkX are different according to the magnetic field direction; 
the SkX is stabilized under the out-of-plane DM interaction for the in-plane magnetic field, while it is stabilized by additionally introducing an easy-axis anisotropy for the out-of-plane magnetic field.
We also discuss the possible multiple-$Q$ states other than the skyrmion crystals in the presence of the out-of-plane DM interaction. 
Our result indicates that noncentrosymmetric magnetic materials with the out-of-plane DM interaction are potential candidates to host the skyrmion crystals. 
\end{abstract}
\maketitle

\section{Introduction}

Noncopalanr spin textures have been intensively studied in condensed matter physics, as they exhibit fascinating topologically-nontrivial states, low-energy excitations, and physical phenomena. 
One of the most typical physical phenomena under noncoplanar magnetic orderings is the topological Hall effect, where a Hall voltage is proportional to neither an applied external field nor a net magnetization. 
This phenomenon has been ubiquitously found in various noncoplanar magnetic orderings irrespective of the details of the spin configurations once the spin texture accompanies a net scalar chirality defined as the triple product of spins even when the effect of relativistic spin-orbit coupling is neglected~\cite{Ohgushi_PhysRevB.62.R6065, Shindou_PhysRevLett.87.116801,tatara2002chirality, Martin_PhysRevLett.101.156402, Akagi_JPSJ.79.083711,hayami_PhysRevB.91.075104, Hayami_PhysRevB.94.024424, Hamamoto_PhysRevB.92.115417,nakazawa2018topological, Matsui_PhysRevB.104.174432}. 
In addition, the noncoplanar spin textures often lead to symmetry lowering compared to the collinear one, which gives rise to a variety of multiferroic responses~\cite{seki2012observation,white2012electric,okamura2013microwave,Mochizuki_PhysRevB.87.134403,tokura2014multiferroics,mochizuki2015dynamical,Christensen_PhysRevX.8.041022,Gobel_PhysRevB.99.060406,bhowal2022magnetoelectric,hayami2022helicity} and nonreciprocal transport~\cite{tokura2018nonreciprocal,Hoshino_PhysRevB.97.024413,ishizuka2020anomalous,hayami2021phase,comment_Hayami}. 
Thus, exploring noncoplanar magnetic orderings has been one of the central issues in both theory and experiment. 

Noncoplanar spin textures are often described by a superposition of multiple spin density waves, which are so-called the multiple-$Q$ states. 
For example, a superposition of three proper-screw spiral waves whose ordering vector lies in a two-dimensional plane and makes a 120$^{\circ}$ angle with each other gives rise to a triangular lattice of a magnetic skyrmion crystal (SkX) or a meron-antimeron crystal depending on a relative phase between the spiral waves. 
Similarly, one constructs a hedgehog lattice by superposing three proper-screw spiral waves with three-dimensional orthogonal ordering vectors. 
Remarkably, such multiple-$Q$ states with noncoplanar magnetic structures have recently been observed in various materials~\cite{nagaosa2013topological, Tokura_doi:10.1021/acs.chemrev.0c00297}: 
the triangular-lattice SkXs in MnSi~\cite{ishikawa1976helical,Muhlbauer_2009skyrmion,Neubauer_PhysRevLett.102.186602}, Fe$_{1-x}$Co$_x$Si~\cite{yu2010real,beille1983long}, FeGe~\cite{yu2011near,lebech1989magnetic}, Cu$_2$OSeO$_3$~\cite{seki2012observation,Adams2012,Seki_PhysRevB.85.220406}, Gd$_2$PdSi$_3$~\cite{Saha_PhysRevB.60.12162,kurumaji2019skyrmion,sampathkumaran2019report,Hirschberger_PhysRevB.101.220401,Kumar_PhysRevB.101.144440,spachmann2021magnetoelastic}, Gd$_3$Ru$_4$Al$_{12}$~\cite{chandragiri2016magnetic,Nakamura_PhysRevB.98.054410,hirschberger2019skyrmion,Hirschberger_10.1088/1367-2630/abdef9}, and EuPtSi~\cite{kakihana2018giant,kaneko2019unique,tabata2019magnetic,kakihana2019unique}, the square-lattice SkXs in GdRu$_2$Si$_2$~\cite{khanh2020nanometric,Yasui2020imaging,khanh2022zoology} and EuAl$_4$~\cite{Shang_PhysRevB.103.L020405,kaneko2021charge,Zhu2022,takagi2022square}, the meron-antimeron crystal in Co$_8$Zn$_9$Mn$_3$~\cite{yu2018transformation}, 
the hedgehog lattice in MnSi$_{1-x}$Ge$_{x}$~\cite{tanigaki2015real,kanazawa2017noncentrosymmetric,fujishiro2019topological,Kanazawa_PhysRevLett.125.137202} and SrFeO$_3$~\cite{Ishiwata_PhysRevB.84.054427,Ishiwata_PhysRevB.101.134406,Rogge_PhysRevMaterials.3.084404,Onose_PhysRevMaterials.4.114420}, and the vortex crystal in Y$_3$Co$_8$Sn$_4$~\cite{takagi2018multiple}.

According to the findings of various types of multiple-$Q$ states in materials, their stabilization mechanisms have been extensively studied from the theoretical side. 
In all the cases, the essence lies in the competing interactions so as to cause a finite-$Q$ spin modulation and its superposition~\cite{hayami2021topological}. 
The most fundamental mechanism is the competition between the ferromagnetic exchange interaction and the Dzyaloshinskii-Moriya (DM) interaction~\cite{dzyaloshinsky1958thermodynamic,moriya1960anisotropic} in addition to an external magnetic field in noncentrosymmetric magnets~\cite{rossler2006spontaneous,Binz_PhysRevLett.96.207202,Binz_PhysRevB.74.214408,Yi_PhysRevB.80.054416,Butenko_PhysRevB.82.052403,Wilson_PhysRevB.89.094411,Mochizuki_PhysRevLett.108.017601}. 
Subsequently, it was demonstrated that the other types of competing interactions, which arise from the short-range spin interactions in Mott insulators~\cite{Momoi_PhysRevLett.79.2081,Okubo_PhysRevLett.108.017206, Kamiya_PhysRevX.4.011023,leonov2015multiply, Lin_PhysRevB.93.064430, Hayami_PhysRevB.93.184413, Hayami_PhysRevB.94.174420, batista2016frustration,Aoyama_PhysRevB.105.L100407}, long-range spin interactions in metals~\cite{Akagi_PhysRevLett.108.096401, Solenov_PhysRevLett.108.096403, Hayami_PhysRevB.90.060402, Ozawa_doi:10.7566/JPSJ.85.103703,Ozawa_PhysRevLett.118.147205, Hayami_PhysRevB.95.224424, Hayami_PhysRevB.99.094420, Wang_PhysRevLett.124.207201, Eto_PhysRevB.104.104425,wang2021skyrmion, Mitsumoto_PhysRevB.105.094427,Mitsumoto_PhysRevB.104.184432,kato2022magnetic,eto2022low}, anisotropic exchange interactions~\cite{Michael_PhysRevB.91.155135,Rousochatzakis2016,amoroso2020spontaneous,yambe2021skyrmion, Hayami_PhysRevB.103.024439, Hayami_PhysRevB.103.054422, Utesov_PhysRevB.103.064414, Wang_PhysRevB.103.104408, amoroso2021tuning,yambe2022effective}, and layer-dependent DM interactions~\cite{Hayami_PhysRevB.105.014408,lin2021skyrmion,hayami2205.03754}, also induce the instabilities toward the multiple-$Q$ states even in centrosymmetric magnets. 
By combining these factors, one obtains a variety of topological spin crystals, such as the SkX~\cite{heinze2011spontaneous,Hayami_doi:10.7566/JPSJ.89.103702,hayami2021field,Kathyat_PhysRevB.103.035111,hayami2022multiple,hayami2205.11272}, meron-antimeron crystal~\cite{Hayami_PhysRevLett.121.137202,Hayami_PhysRevB.103.024439,Wang_PhysRevB.103.104408,Hayami_PhysRevB.103.054422,Hayami_PhysRevB.104.094425}, and hedgehog lattice~\cite{grytsiuk2020topological,Okumura_PhysRevB.101.144416,Shimizu_PhysRevB.103.054427,Aoyama_PhysRevB.103.014406,Mendive-Tapia_PhysRevB.103.024410,Kato_PhysRevB.104.224405} as the ground state.

In the present study, we revisit the effect of the DM interaction on the stabilization of the SkX. 
Although the DM interaction occurs in noncentrosymmetric lattice systems, its behavior is largely dependent on point groups, which gives rise to the SkXs with different helicity and vorticity around the core. 
For example, the DM interaction favors the N\'eel-type (Bloch-type) SkX under the point groups $C_{4v}$ and $C_{6v}$ ($D_4$, $D_6$, $T$, and $O$), while it favors the anti-type SkX under the point groups $D_{2d}$ and $S_4$. 
The contribution of the DM interaction to the free energy in each point group is given as Lifshitz invariants~\cite{dzyaloshinskii1964theory,kataoka1981helical,Bogdanov89,Bogdanov94}. 
On the other hand, there are several point groups ($D_{3h}$, $C_{3h}$, and $T_d$) that are not included in the above situation, where the role of the DM interaction on the SkXs has not been fully clarified yet. 

Among them, we investigate the instability toward the SkX in the spin system with the out-of-plane DM vector under the point group $D_{3h}$. 
By constructing an effective spin model with the out-of-plane DM interaction in momentum space and performing the simulated annealing for the model, we examine the tendency of the SkX in a wide range of model parameters. 
We find a minimal condition to stabilize the SkX under an external magnetic field along the in-plane and out-of-plane directions; the SkX is stabilized by considering the Heisenberg-type exchange interaction and the out-of-plane DM interaction for the in-plane external magnetic field, while it is stabilized by additionally introducing an easy-axis anisotropy for the out-of-plane magnetic field. 
For the latter, we obtain two types of the SkXs, which are transformed by changing the magnitude of the magnetic field. 
We also discuss the instabilities toward the other multiple-$Q$ states under the $D_{3h}$ system. 
Our results indicate a possibility of the SkX based on the DM-interaction mechanism that is distinct from the case in the polar, chiral, and ($D_{2d}, S_4$) systems. 

The paper is arranged as follows: 
In Sec.~\ref{sec: Model and method}, we show an effective spin model with the out-of-plane DM interaction on a triangular lattice belonging to the point group $D_{3h}$ and outline the numerical method based on the simulated annealing. 
We discuss the instability toward the SkX in the presence of the out-of-plane DM interaction when applying the external magnetic field along the in-plane and out-of-plane directions in Sec.~\ref{sec: Results}. 
Finally, in Sec.~\ref{sec: Summary}, we briefly summarize the results of the paper.
In Appendix~\ref{sec: Spin configurations under strong easy-axis anisotropy}, we present the spin configurations in real and momentum spaces for large single-ion anisotropy.

\section{Model and method}
\label{sec: Model and method}

\begin{figure}[t!]
\begin{center}
\includegraphics[width=1.0 \hsize ]{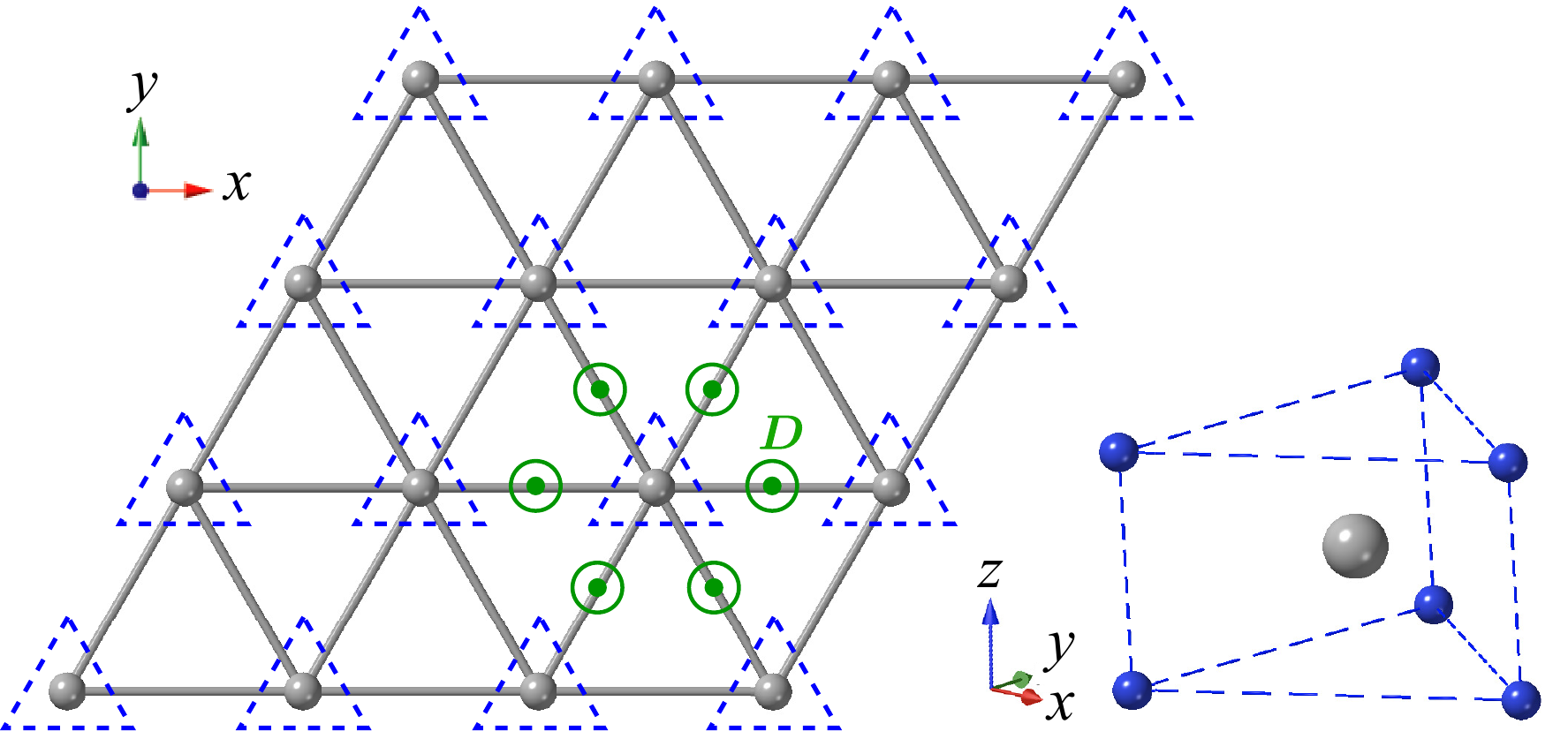} 
\caption{
\label{fig: lattice}
Triangular lattice structure under the point group $D_{3h}$. 
Each lattice site is surrounded by the ligand ions satisfying $D_{3h}$ symmetry shown in the right-bottom panel.
The out-of-plane DM vectors are also shown.  
}
\end{center}
\end{figure}

We consider a spin model on a two-dimensional triangular lattice under the point group $D_{3h}$ in Fig.~\ref{fig: lattice}; the triangular-lattice plane lies in the $xy$ plane and the lattice constant is set as unity. 
We suppose that site symmetry at each lattice site is $D_{3h}$ by implicitly considering a triangular-prism alignment of the ligand ions, as shown in the right-bottom panel of Fig.~\ref{fig: lattice}. 
A similar alignment of the ligand ion has been found in the van der Waals magnets~\cite{doi:10.1063/5.0039979}, such as the transition metal dichalcogenides~\cite{puotinen1961crystal,manzeli20172d} and Fe$_3$GeTe$_2$~\cite{deiseroth2006fe3gete2,Park_PhysRevB.103.104410,ding2019observation}.
There is no inversion center at any bonds, while the mirror symmetry with respect to the $xy$ plane is present. 
Then, the DM vector at any bonds has the only out-of-plane component, as shown in Fig.~\ref{fig: lattice}.

A minimum spin model to incorporate the effect of such a DM interaction is given by 
\begin{align}
\mathcal{H}^{\rm real-space}=&  -\sum_{i,j} \left[ J_{ij} \bm{S}_{i} \cdot \bm{S}_{j}+ 
D^z_{ij} (\bm{S}_{i} \times \bm{S}_{j})^{z} \right], \nonumber
\end{align}
where $\bm{S}_i=(S_i^x, S_i^y, S_i^z)$ represents the classical spin at site $i$ with the fixed length $|\bm{S}_i|=1$. 
The first term corresponds to the Heisenberg- exchange interaction and the second term represents the DM exchange interaction. 
The latter DM interaction microscopically originates from the spin-dependent hopping (the atomic spin-orbit coupling) like the Rashba and Dresselhaus systems, although the spin component is different from each other~\cite{Banerjee_PhysRevX.4.031045,Kathyat_PhysRevB.102.075106,Mukherjee_PhysRevB.103.134424}:  In the case of the Rashba (Dresselhaus) case, the in-plane component of the DM vector becomes nonzero, while 
only the out-of-plane one becomes nonzero irrespective of the bond in the case of the present system.

By performing the Fourier transformation, an effective spin model with the momentum-resolved interaction is given by 
\begin{align}
\mathcal{H} = -\sum_{\bm{q}}
\left[J_{\bm{q}} \bm{S}_{\bm{q}} \cdot  \bm{S}_{-\bm{q}}+i D^z_{\bm{q} }\left({\bm S}_{\bm{q}}\times{\bm S}_{-\bm{q}}\right)^{z}
\right], 
\label{eq: Ham}
\end{align}
where $\bm{S}_{\bm{q}}$ is the Fourier transform of the classical localized spin; $\bm{q}$ is the wave vector. 
The first term represents the Heisenberg-type exchange interaction and the second term represents the DM-type exchange interaction; $J_{\bm{q}}$ and $D_{\bm{q}}$ correspond to the Fourier transform of $J_{ij}$ and $D^z_{ij}$, respectively. 
In order to focus on the role of the out-of-plane DM interaction on magnetic orderings, we neglect bond-dependent exchange interactions arrowed by the point group $D_{3h}$.

The $\bm{q}$ dependences of $J_{\bm{q}}$ and $D_{\bm{q}}$ are determined by the spin interactions in real space. 
When analyzing a low-temperature phase diagram, the model is simplified by extracting the dominant interactions in Eq.~(\ref{eq: Ham}). 
We assume that the dominant interactions are dictated by the specific wave vectors along the high-symmetric lines $\pm \bm{Q}_1=\pm (Q,0)$, $\pm \bm{Q}_2=\pm (-Q/2,\sqrt{3}Q/2)$, and $\pm \bm{Q}_3=\pm (-Q/2,-\sqrt{3}Q/2)$ with $Q=\pi/3$, which are not located at $\bm{q}=0$ and the Brilloune zone boundary; $\bm{Q}_1$, $\bm{Q}_2$, and $\bm{Q}_3$ are connected by the threefold rotational symmetry of the triangular lattice, which imposes on $J_{\bm{Q}_1}=J_{\bm{Q}_2}=J_{\bm{Q}_3}$ and $D_{\bm{Q}_1}=D_{\bm{Q}_2}=D_{\bm{Q}_3}$. 
It is noted that the following results are not qualitatively altered for the other choice of the ordering vectors. 
This position of the ordering vector can be obtained by considering the appropriate $J_{ij}$ and $D^z_{ij}$, although we give it phenomenologically. 
Then, the model in Eq.~(\ref{eq: Ham}) reduces to 
\begin{align}
\mathcal{H} = -\sum_{\nu}
\left[J \bm{S}_{\bm{Q}_\nu} \cdot  \bm{S}_{-\bm{Q}_\nu}+i D \left({\bm S}_{\bm{Q}_\nu}\times{\bm S}_{-\bm{Q}_\nu}\right)^z
\right], 
\label{eq: Ham2}
\end{align}
where we omit the subscript of $J_{\bm{Q}_\nu}$ and $D_{\bm{Q}_\nu}$ for notational simplicity. 
It is noted that a spin state with the intensity $\bm{Q}_1$ has the same energy as those with $\bm{Q}_2$ or $\bm{Q}_3$ owing to the threefold rotational symmetry of the Hamiltonian. 
There is no spin anisotropy in the $xy$ plane owing to the absence of the bond-dependent interaction.  

The model in Eq.~(\ref{eq: Ham2}) is regarded as an effective spin model to investigate the ground-state phase diagram in the original spin model of $\mathcal{H}^{\rm real-space}$ for magnetic insulators in the presence of relativistic spin-orbit coupling. 
In this case, the momentum-resolved interaction is related to the Fourier transform of the real-space short-range spin interactions, such as the frustrated exchange interactions. 
Besides, the model in Eq.~(\ref{eq: Ham2}) can be regarded as an effective spin model for magnetic metals belonging to the same point group, where the interactions in Eq.~(\ref{eq: Ham2}) are mapped onto the real-space long-range spin interactions, such as the Ruderman-Kittel-Kasuya-Yosida (RKKY) interaction~\cite{Ruderman, Kasuya, Yosida1957}. 
Although the interactions between the further-neighboring spins play an important role in the RKKY system, only the out-of-plane DM interaction is present due to horizontal mirror symmetry under $D_{3h}$ point group.
The instabilities toward the SkXs based on a similar effective spin model with a particular set of the momentum-resolved interactions under other point groups have been investigated in both insulators~\cite{leonov2015multiply,Hayami_PhysRevB.103.224418,Hayami_PhysRevB.105.014408,hayami2022multiple,hayami2022skyrmion} and metals~\cite{Hayami_PhysRevB.95.224424,Wang_PhysRevLett.124.207201,Hayami_PhysRevB.103.024439,Hayami_PhysRevB.103.054422,yambe2021skyrmion,yambe2022effective}. 
Hereafter, we set $J=1$ as the energy unit and change $D$ as a phenomenological parameter. 

To enhance the instability toward the SkX under the out-of-plane DM interaction, we additionally introduce the local Hamiltonian including the effect of an external magnetic field. 
We consider two different directions of the magnetic field: 
One is the in-plane magnetic field, whose Hamiltonian is given as the Zeeman coupling by 
\begin{align}
\mathcal{H}^{\rm local}_x = -H^x \sum_{i}  S^x_i. 
\label{eq: Ham_localHx}
\end{align}
The other is the out-of-plane magnetic field, which is given by 
\begin{align}
\mathcal{H}^{\rm local}_z = -A\sum_{i}(S^z_i)^2 -H^z \sum_{i}  S^z_i, 
\label{eq: Ham_localHz}
\end{align}
where we consider the effect of the easy-axis single-ion anisotropy $A>0$. 
We take $A$, $H^x$, and $H^z$ as phenomenological parameters in addition to $D$. 

The low-temperature magnetic phase diagrams in the model consisting of Eqs.~(\ref{eq: Ham2}), (\ref{eq: Ham_localHx}), and (\ref{eq: Ham_localHz}) are constructed by performing simulated annealing for the system size with a cluster of $N=96^2$ spins.  
The simulations are carried out based on the standard Metropolis local updates in real space. 
In each model parameter set of $D$, $A$, $H^x$, and $H^z$, we adopt a different random spin configuration and start the simulations from high temperatures, which are typically set as $T_0=1$-$10$. 
By gradually reducing the temperature of the system by the rate $\alpha=0.99995$-$0.999995$, we try to find the optimal spin configurations for given model parameters. 
The target temperature is set as 0.01. 
Once the temperature reaches the target temperature, we perform $10^5$-$10^6$ Monte Carlo sweeps for measurements after equilibration. 
In the vicinity of the phase boundaries where several different spin configurations are obtained in the simulations, we also start the simulations from the spin configurations obtained at low temperatures to determine the phase boundary. 

The numerically-obtained spin configurations are distinguished by the spin- and chirality-related quantities. 
In the spin sector, we compute the $\mu=x,y,z$ component of the magnetic moment with wave vector $\bm{q}$, whose expression is given by 
\begin{align}
\label{eq:mq}
m^\mu_{\bm{q}} &= \sqrt{\frac{S^{\mu}_s(\bm{q})}{N}}, \\
S^{\mu}_s(\bm{q}) &= \frac1N \sum_{j,l}  S_j^{\mu} S_l^{\mu}  e^{i \bm{q} \cdot (\bm{r}_j-\bm{r}_l)},  
\end{align}
where $S^{\mu}_s$ is the spin structure factor; $\bm{r}_j$ is the position vector at site $j$. 
We also compute $S^{yz}_{s} (\bm{q})=S^{y}_{s} (\bm{q})+S^{z}_{s} (\bm{q})$, $S^{xy}_{s} (\bm{q})=S^{x}_{s} (\bm{q})+S^{y}_{s} (\bm{q})$, and $(m^{xy}_{\bm{q}})^2=(m^{x}_{\bm{q}})^2+(m^{y}_{\bm{q}})^2$. 
The uniform magnetization is calculated by $M^{\mu}=(1/N)\sum_i S^\mu_i$, which corresponds to the $\bm{q}=\bm{0}$ component of $m^\alpha_{\bm{q}}$; we also define the in-plane magnetization $M^{xy}=\sqrt{(M^x)^2+(M^y)^2}$. 

In the chirality sector, we calculate the spin scalar chirality defined by the triple scalar product of neighboring spins on a triangle plaquette, which is given by 
\begin{align}
\chi_0=\frac{1}{N}\sum_{\bm{R}} \bm{S}_{j}\cdot (\bm{S}_k \times \bm{S}_l), 
\end{align}
where $\bm{R}$ denotes the position vectors at the centers of the upward and downward triangles consisting of spins $\bm{S}_j$, $\bm{S}_k$, and $\bm{S}_l$ in counterclockwise order. 
The nonzero $\chi_0$ indicates the emergence of the topological spin textures, such as the SkX.

\section{Results}
\label{sec: Results}

We show the SkX instability in the presence of the out-of-plane DM interaction in two situations. 
One is the situation under the in-plane magnetic field in Sec.~\ref{sec: In an in-plane magnetic field}, and the other is that under the out-of-plane magnetic field in Sec.~\ref{sec: In an out-of-plane magnetic field}. 

\subsection{In an in-plane magnetic field}
\label{sec: In an in-plane magnetic field}

\begin{figure}[t!]
\begin{center}
\includegraphics[width=1.0 \hsize ]{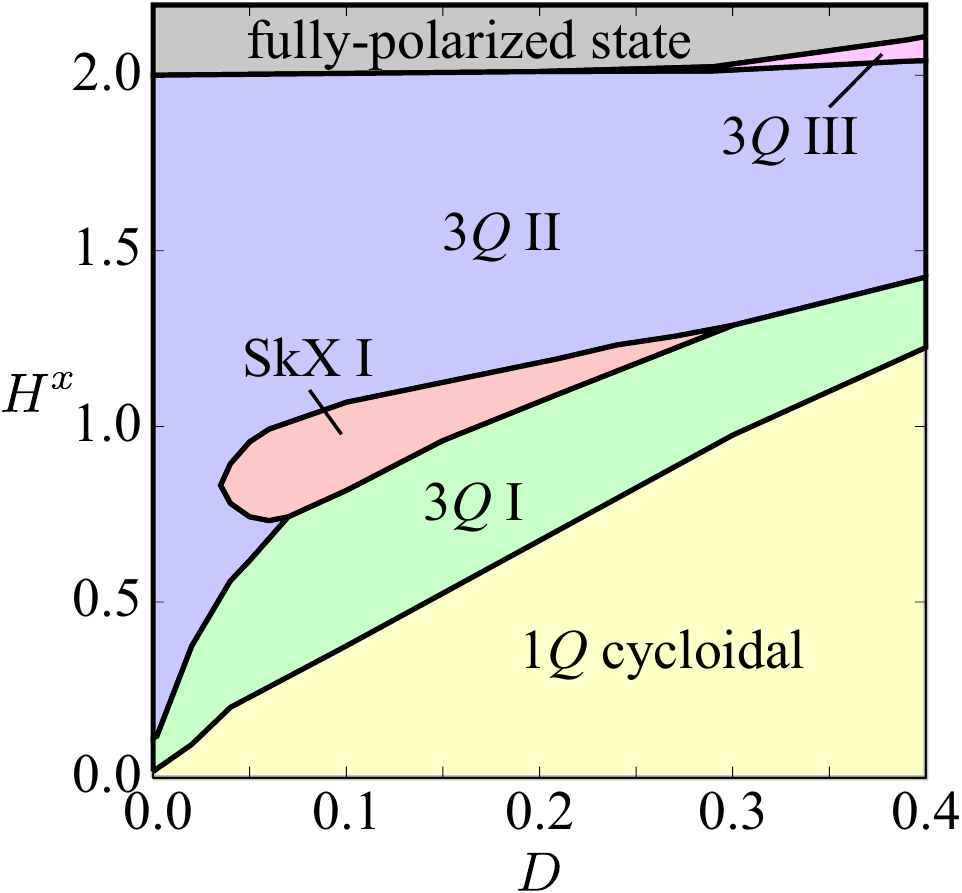} 
\caption{
\label{fig: PD_D-Hx}
Magnetic phase diagram of the model Hamiltonian $\mathcal{H}+\mathcal{H}^{\rm local}_x$ in Eqs.~(\ref{eq: Ham2}) and (\ref{eq: Ham_localHx}) in the plane of $D$ and $H^x$ at $H^z=A=0$ obtained by simulated annealing at $T=0.01$. 
The spin and scalar chirality configurations in each phase are shown in Fig.~\ref{fig: spin_D-Hx}. 
}
\end{center}
\end{figure}

\begin{figure*}[t!]
\begin{center}
\includegraphics[width=1.0 \hsize ]{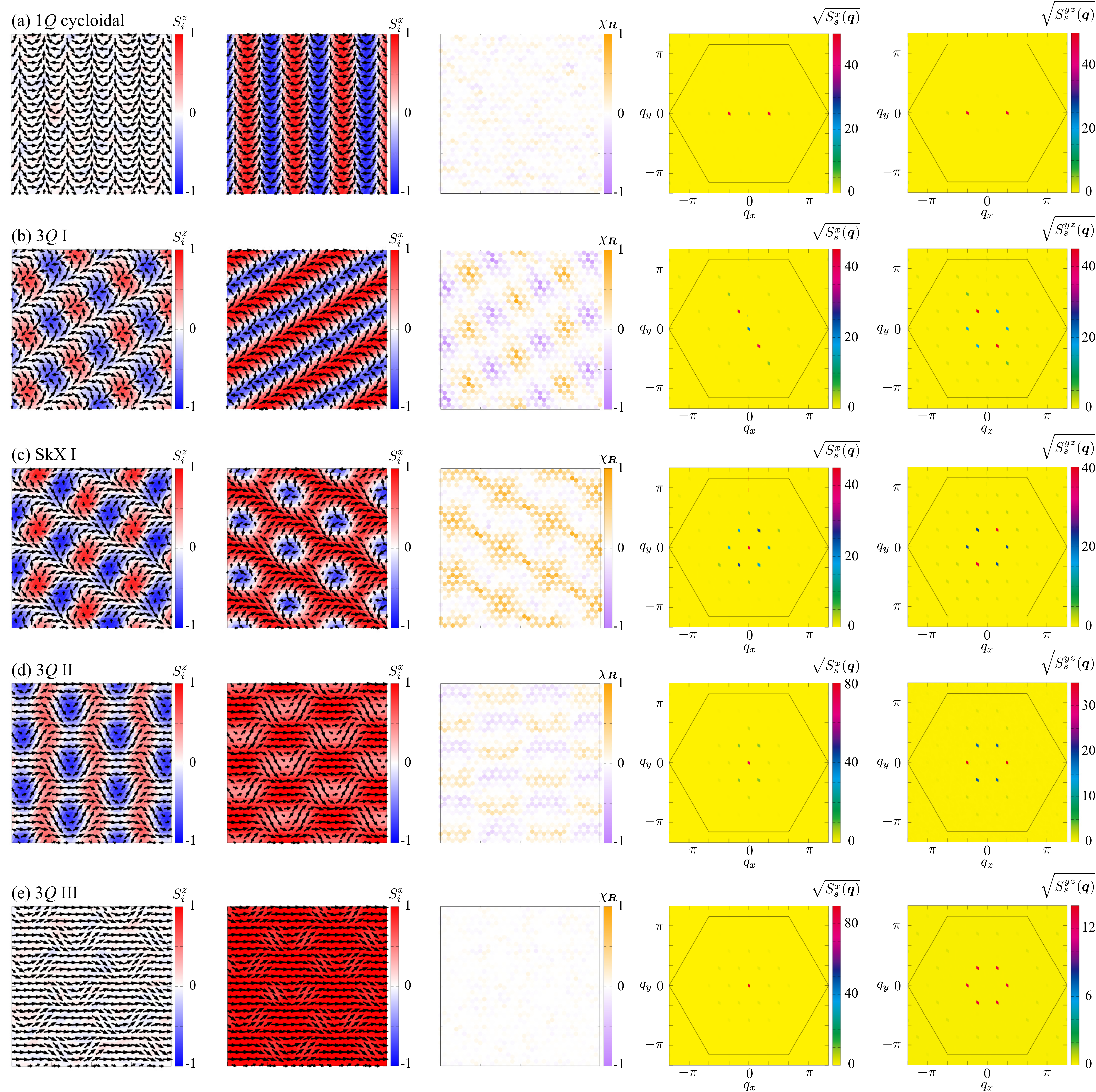} 
\caption{
\label{fig: spin_D-Hx}
(Left panel) Snapshots of the spin configurations in (a) the single-$Q$ (1$Q$) cycloidal spiral state for $D=0.1$ and $H^x=0.2$, (b) triple-$Q$ (3$Q$) I state for $D=0.1$ and $H^x=0.7$, (c) SkX I for $D=0.1$ and $H^x=1$, (d) 3$Q$ II state for $D=0.1$ and $H^x=1.5$, and (e) 3$Q$ III state for $D=0.35$ and $H^x=2.05$. 
The contours show the $z$ component of the spin moment, while the arrows represent the $xy$ components of the spin moment. 
(Second left panel) Contour plots of $S_i^x$ corresponding to the left panel. 
(Middle panel) Snapshots of the scalar chirality configurations; $\chi_{\bm{R}}=\bm{S}_{j}\cdot (\bm{S}_k \times \bm{S}_l)$. 
(Second right and right panels) The square root of the $x$ and $yz$ components of the spin structure factor, respectively. 
The hexagons with a solid line in the right two columns show the first Brillouin zone.
}
\end{center}
\end{figure*}

\begin{figure}[t!]
\begin{center}
\includegraphics[width=0.85 \hsize ]{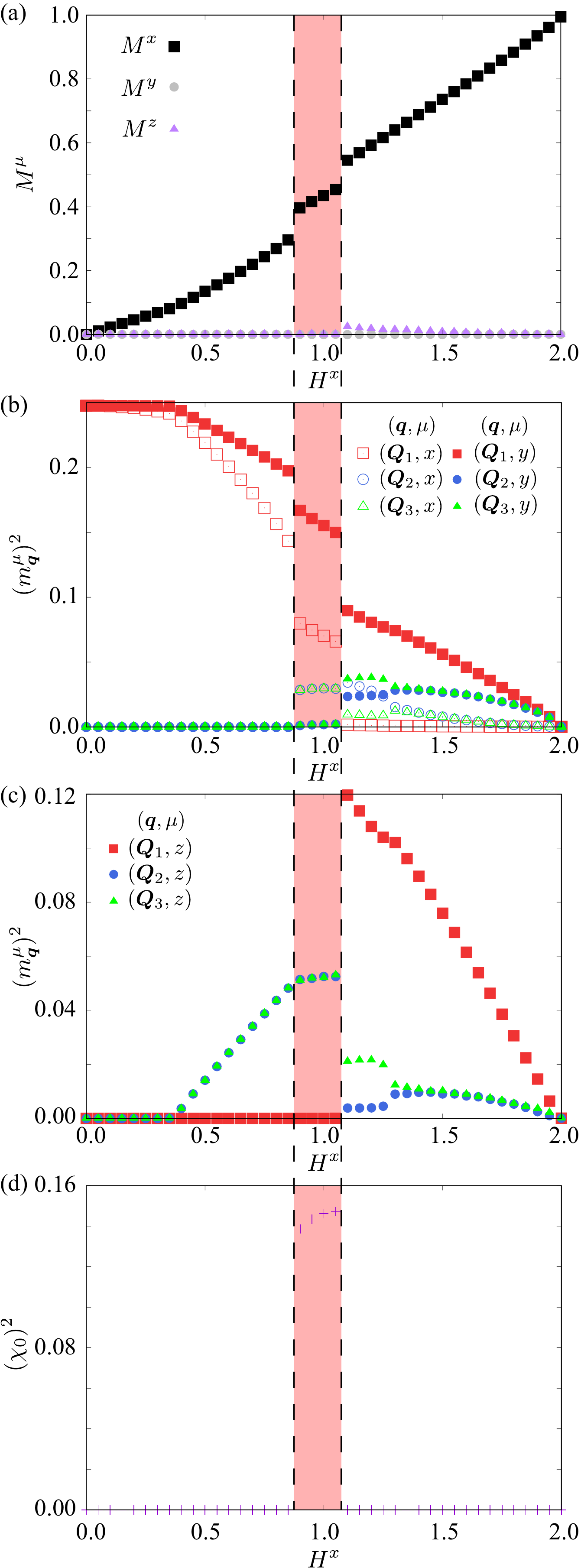} 
\caption{
\label{fig: Mq_D-Hx}
$H^x$ dependences of (a) $M^x$, $M^y$, and $M^z$, (b) $(m^{\mu}_{\bm{q}})^2$ for $\mu=x,y$ and $\bm{q}=\bm{Q}_1, \bm{Q}_2, \bm{Q}_3$, (c) $(m^{z}_{\bm{q}})^2$, and (d) $(\chi_0)^2$ at $D=0.1$. 
The vertical dashed lines represent the boundaries between the SkX I and the other magnetic phases.
}
\end{center}
\end{figure}

We examine the instability toward the SkX under the in-plane field by considering the model Hamiltonian $\mathcal{H}+\mathcal{H}^{\rm local}_x$ in Eqs.~(\ref{eq: Ham2}) and (\ref{eq: Ham_localHx}), i.e., $A=H^z=0$. 
Figure~\ref{fig: PD_D-Hx} shows the magnetic phase diagram while changing $D$ and $H^x$ at low temperature $T=0.01$. 
In the phase diagram, there are five phases except for the fully-polarized state where the spins are aligned in the field direction. 
The spin and chirality configurations in real space as well as the spin structure factor for each phase are shown in Fig.~\ref{fig: spin_D-Hx}. 
We also show the $H^x$ dependences of spin- and chirality-related quantities at $D=0.1$ in Fig.~\ref{fig: Mq_D-Hx}. 
When the optimal spin configuration is characterized by $\bm{m}_{\bm{Q}_1} \neq 0$, that characterized by $\bm{m}_{\bm{Q}_2} \neq 0$ or $\bm{m}_{\bm{Q}_3} \neq 0$ are also obtained depending on initial spin configurations.
Then, we appropriately sort $\bm{m}_{\bm{Q}_\nu}$ in Fig.~\ref{fig: Mq_D-Hx} for better readability. 
The same situation also holds for other multiple-$Q$ states. 

At zero field ($H^x=0$), the single-$Q$ cycloidal spiral state is stabilized for any $D$, as shown in Fig.~\ref{fig: PD_D-Hx}. 
As shown in the real-space spin configuration and the spin structure factor in Fig.~\ref{fig: spin_D-Hx}(a), the spin state is characterized by the spiral modulation in the $xy$ plane so as to gain the energy by the out-of-plane DM interaction. 
The magnitude of $m^x_{\bm{Q}_1}$ and $m^y_{\bm{Q}_1}$ is the same, as shown in Fig.~\ref{fig: Mq_D-Hx}(b). 
When introducing the in-plane magnetic field ($H^x \neq 0$), the $x$ component of the uniform magnetization is induced in Fig.~\ref{fig: Mq_D-Hx}(a). 
Simultaneously, the $y$ component of spins at $\bm{Q}_1$ becomes larger than the $x$ component so as to gain the energy by the in-plane magnetic field, as shown in Fig.~\ref{fig: Mq_D-Hx}(b). 
Accordingly, there are intensities at the higher-harmonic wave vector $2\bm{Q}_1$ shown in the right two panels in Fig.~\ref{fig: spin_D-Hx}(a). 
Meanwhile, there is no scalar chirality in this state due to the coplanar spin structure, as shown in the middle panel of Fig.~\ref{fig: spin_D-Hx}(a) and Fig.~\ref{fig: Mq_D-Hx}(d). 

While increasing $H^x$, the single-$Q$ cycloidal spiral state turns into the other state with a different spin configuration; the single-$Q$ spin configuration with $m^{x}_{\bm{Q}_1}$ and $m^{y}_{\bm{Q}_1}$ is gradually modulated so as to have $m^{z}_{\bm{Q}_2}$ and $m^{z}_{\bm{Q}_3}$, as shown in Fig.~\ref{fig: spin_D-Hx}(c). 
Thus, the dominant component of the spin configuration is the cycloidal spiral component along the $\bm{Q}_1$ direction, i.e, $m^{x}_{\bm{Q}_1},m^{y}_{\bm{Q}_1} > m^{z}_{\bm{Q}_2},m^{z}_{\bm{Q}_3}$. 
Such a behavior is also found in the spin structure factor in the right two panels of Fig.~\ref{fig: spin_D-Hx}(b). 
The appearance of the intensities at the other $\bm{Q}_\nu$ indicates the triple-$Q$ formation, where we denote this state as the triple-$Q$ I state. 
In contrast to the single-$Q$ cycloidal state, the spin configuration in this state is noncoplanar, which results in the local scalar chirality distribution $\chi_{\bm{R}}$, as shown in the middle panel in Fig.~\ref{fig: spin_D-Hx}(b). 
The modulation of $\chi_{\bm{R}}$ is characterized by the chirality density wave along the $\bm{Q}_2$ and $\bm{Q}_3$ directions. 
The correspondence of the chirality density wave with $m^{z}_{\bm{Q}_2}$ and $m^{z}_{\bm{Q}_3}$ is understood from the real-space picture where the sign of the local scalar chirality corresponds to that of $S_i^z$, as shown in the left panel of Fig.~\ref{fig: spin_D-Hx}(b). 
It is noted that there is no net component of the scalar chirality in the triple-$Q$ I state in Fig.~\ref{fig: Mq_D-Hx}(d).

The transition from the single-$Q$ cycloidal state to the triple-$Q$ I state is understood from the competition of the out-of-plane DM interaction and the Zeeman energy: The former tends to lead to the in-plane spiral, while the latter favors the $yz$-plane spin modulation, which is perpendicular to the magnetic field. 
In this sense, the magnetic field destabilizes the single-$Q$ cycloidal state, which leads to the phase transition to the spin state with the additional modulations perpendicular to the field. 
The reason why the $z$ component of spins at $\bm{Q}_2$ and $\bm{Q}_3$ takes a finite value rather than the $y$ component is to reduce the exchange energy loss that arises from the normalization of the spin length. 
In other words, when there is a $y$-spin component at $\bm{Q}_1$, the $z$ component of spins is favored at $\bm{Q}_2$ and $\bm{Q}_3$ to reduce the higher harmonic components of $\bm{S}_{\bm{q}}$. 
Accordingly, the phase boundary between the single-$Q$ cycloidal state and the triple-$Q$ I state moves toward large $H^x$ while increasing $D$ since the DM interaction tends to favor the in-plane spin configuration, as mentioned above.

In the intermediate $D$ region for $0.04 \lesssim D \lesssim 0.3$, the triple-$Q$ I state turns into the SkX I with the jumps of $M^x$, $(m^x_{\bm{Q}_\nu})^2$, $(m^y_{\bm{Q}_\nu})^2$, and $(\chi_{0})^2$, as shown in Figs.~\ref{fig: Mq_D-Hx}(a), \ref{fig: Mq_D-Hx}(b), and \ref{fig: Mq_D-Hx}(d), respectively. 
Compared to the triple-$Q$ I state, $(m^x_{\bm{Q}_{2,3}})^2$ and $(m^y_{\bm{Q}_{2,3}})^2$ become nonzero in the SkX I state, while $(m^z_{\bm{Q}_\nu})^2$ does not change qualitatively. 
One finds similar real-space $z$-spin profiles but different $xy$-spin ones between the triple-$Q$ I state and SkX I, as shown in the left and second left panels of Figs.~\ref{fig: spin_D-Hx}(b) and \ref{fig: spin_D-Hx}(c). 
The feature of the SkX is found in the $x$-spin component parallel to the field direction shown in the second left panel of Fig.~\ref{fig: spin_D-Hx}(c), where the skyrmion cores located at $S_i^x \simeq -1$ constitute a slightly deformed triangular lattice. 
Accordingly, the SkX I accompanies a net scalar chirality, as shown in the middle panel of Fig.~\ref{fig: spin_D-Hx}(c) [see also Fig.~\ref{fig: Mq_D-Hx}(d)]. 
Although the net scalar chirality takes a positive value in the case of Fig.~\ref{fig: spin_D-Hx}(c), it becomes negative by reversing $S_i^z$ without the energy change; the states with $\pm \chi_0$ are degenerate. 
Thus, the topological (skyrmion) number in this state becomes $\pm 1$. 

The emergence of the SkX I is also owing to the competition between the out-of-plane DM interaction and the in-plane magnetic field: 
The former favors the in-plane cycloidal spiral state, while the latter favors the $yz$-plane spiral state. 
The present SkX I is regarded as a coexisting state of in-plane cycloidal and out-of-plane  spiral states; the spin configuration is characterized by a superposition of the in-plane cycloidal spiral along the $\bm{Q}_1$ direction and the out-of-plane spiral along the $\bm{Q}_2$ and $\bm{Q}_3$ directions with $m^z_{\bm{Q}_\nu}>m^x_{\bm{Q}_\nu}>m^y_{\bm{Q}_\nu}$ for $\nu=2,3$ in Figs.~\ref{fig: Mq_D-Hx}(b) and \ref{fig: Mq_D-Hx}(c). 
So, the SkX I has additional in-plane intensities at $\bm{Q}_2$ and $\bm{Q}_3$ compared to 3$Q$ I, where the triple-$Q$ peaks of $x$ component appear to induce large $M^x$.
Thus, the SkX I gains energy by both the DM interaction and the magnetic field. 
In other words, the magnetic field direction is important to induce the SkX in the presence of magnetic anisotropy. 
Indeed, there is no longer the SkX I once the magnetic field direction changes from the in-plane to the out-of-plane direction, as discussed below.   
A similar relationship between the magnetic anisotropy and the magnetic field has been discussed in different systems, such as centrosymmetric magnets with the easy-axis or easy-plane anisotropy~\cite{leonov2015multiply, Lin_PhysRevB.93.064430, Hayami_PhysRevB.93.184413,hayami2020multiple,Hayami_PhysRevB.103.224418,hayami2022skyrmion}. 

While increasing $H$ from the SkX I, the spin configuration abruptly changes when entering into a triple-$Q$ II state as shown in Figs.~\ref{fig: Mq_D-Hx}(b) and \ref{fig: Mq_D-Hx}(c) as well as the jumps of $M^x$ and $(\chi_0)^2$ in Figs.~\ref{fig: Mq_D-Hx}(a) and \ref{fig: Mq_D-Hx}(d). 
The main difference is found in the spiral nature of the dominant component of $\bm{m}_{\bm{Q}_\nu}$, which changes from the in-plane cycloidal to the out-of-plane cycloidal spirals. 
This is because the out-of-plane cycloidal spiral has more energy gain by the Zeeman coupling than the in-plane cycloidal spiral. 
This tendency is found in the suppression of $(m^{x}_{\bm{Q}_1})^2$ and the development of $(m^z_{\bm{Q}_1})^2$ in Figs.~\ref{fig: Mq_D-Hx}(b) and \ref{fig: Mq_D-Hx}(c). 
Another difference is a spontaneous magnetization along the $z$ direction, as shown in the right panel of Fig.~\ref{fig: spin_D-Hx}(d) and Fig.~\ref{fig: Mq_D-Hx}(a), which might be attributed to a nonzero triple-$Q$ $z$-spin modulation giving rise to the higher harmonic contribution as $\bm{Q}_1+\bm{Q}_2+\bm{Q}_3=\bm{0}$. 
The magnitudes of $(\bm{m}_{\bm{Q}_2})^2$ and $(\bm{m}_{\bm{Q}_3})^2$ are different for small $H^x$, while those are almost the same for large $H^x$,  as shown in Figs.~\ref{fig: Mq_D-Hx}(b) and \ref{fig: Mq_D-Hx}(c). 
The real-space spin and chirality configurations in the triple-$Q$ II state are shown in the left three panels of Fig.~\ref{fig: spin_D-Hx}(d). 
The scalar chirality is distributed in a checkerboard way owing to the double-$Q$ modulation of $\chi_{\bm{R}}$ with the $\bm{Q}_2$ and $\bm{Q}_3$ components, as shown in the middle panel of Fig.~\ref{fig: spin_D-Hx}(d); there is no net scalar chirality similar to the triple-$Q$ I state [see Fig.~\ref{fig: Mq_D-Hx}(d)]. 

While further increasing $H^x$, the triple-$Q$ II state continuously changes into the fully-polarized state except for large $D$.
For large $D$, another triple-$Q$ state denoted as a triple-$Q$ III state appears in the narrow range of $H^x$, as shown in Fig.~\ref{fig: PD_D-Hx}.  
The spin configuration is characterized by a superposition of the triple-$Q$ in-plane spin state to gain the energy by the DM interaction, where the intensities at $\bm{Q}_1$-$\bm{Q}_3$ are equivalent shown in the right two panels of Fig.~\ref{fig: spin_D-Hx}(e). 
Thus, the spin configuration is coplanar without the local scalar chirality, as found in the left three panels of Fig.~\ref{fig: spin_D-Hx}(e). 
This state also turns into the fully-polarized state with an increase of $H^x$.

\subsection{In an out-of-plane magnetic field}
\label{sec: In an out-of-plane magnetic field}

\begin{figure}[t!]
\begin{center}
\includegraphics[width=1.0 \hsize ]{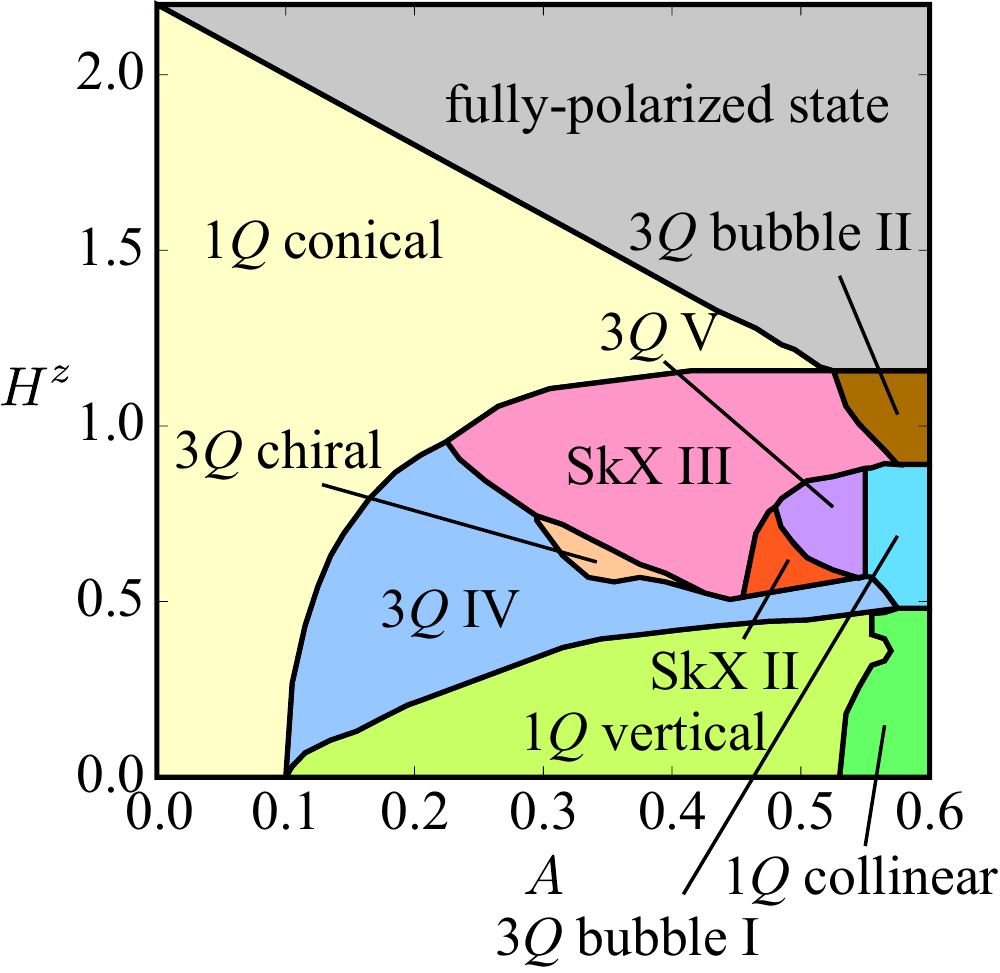} 
\caption{
\label{fig: PD_A-Hz}
Magnetic phase diagram of the model Hamiltonian $\mathcal{H}+\mathcal{H}^{\rm local}_z$ in Eqs.~(\ref{eq: Ham2}) and (\ref{eq: Ham_localHz}) in the plane of $A$ and $H^z$ at $D=0.1$ and $H^x=0$ obtained by simulated annealing at $T=0.01$. 
The spin and scalar chirality configurations in each phase are shown in Figs.~\ref{fig: spin_A-Hz}, \ref{fig: spin_A-Hz_2}, and \ref{fig: spin_A-Hz_app}. 
}
\end{center}
\end{figure}

\begin{figure*}[t!]
\begin{center}
\includegraphics[width=1.0 \hsize ]{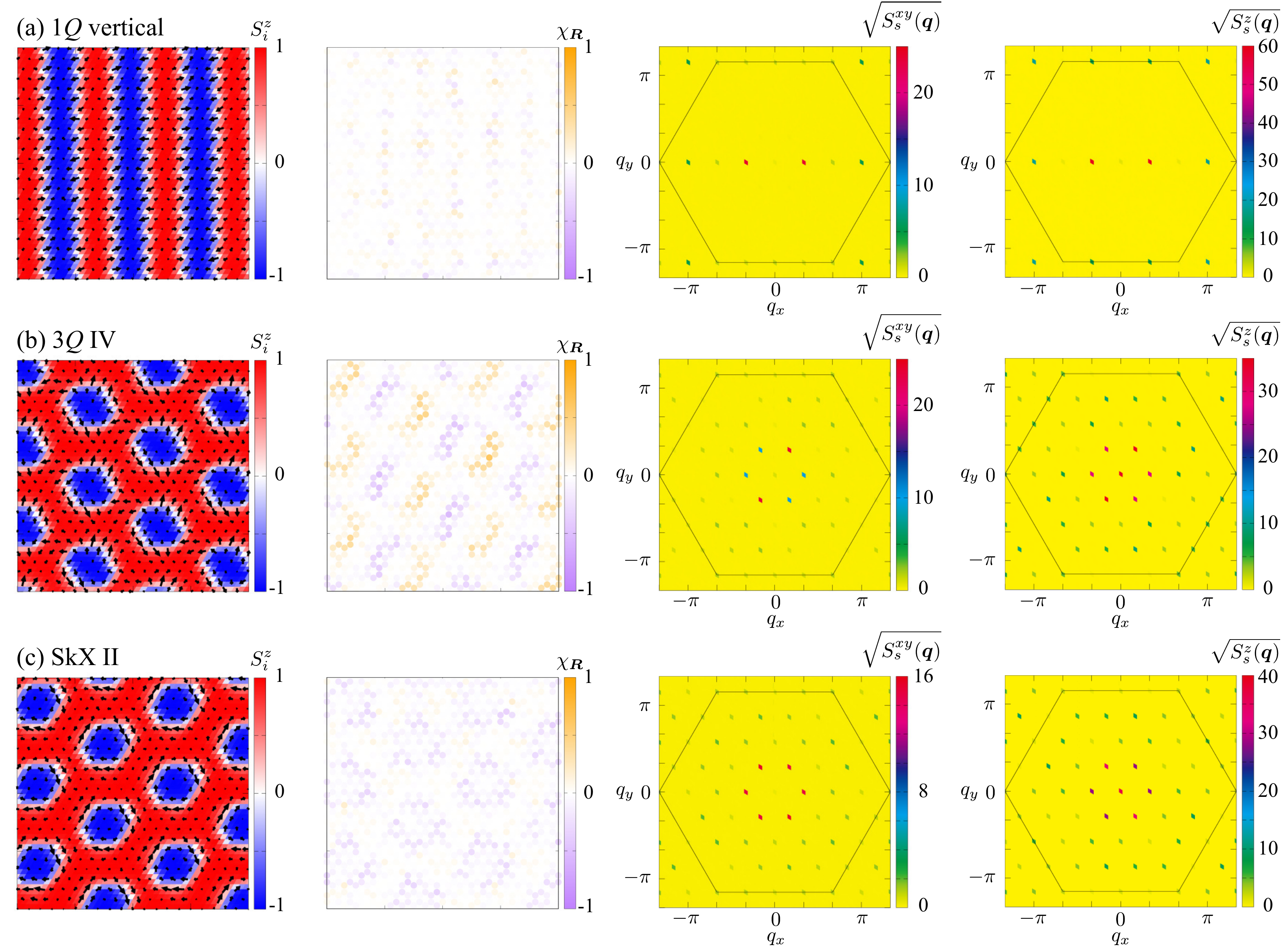} 
\caption{
\label{fig: spin_A-Hz}
(Left panel) Snapshots of the spin configurations in (a) the single-$Q$ (1$Q$) vertical spiral state for $H^z=0.2$, (b) triple-$Q$ (3$Q$) IV state for $H^z=0.5$, and (c) SkX II for $H^z=0.55$ at $A=0.5$. 
The contours show the $z$ component of the spin moment, while the arrows represent the $xy$ components of the spin moment. 
(Middle left panel) Snapshots of the scalar chirality configurations; $\chi_{\bm{R}}=\bm{S}_{j}\cdot (\bm{S}_k \times \bm{S}_l)$. 
(Middle right and right panels) The square root of the $xy$ and $z$ components of the spin structure factor, respectively. 
The hexagons with a solid line in the right two columns show the first Brillouin zone.
}
\end{center}
\end{figure*}

\begin{figure*}[t!]
\begin{center}
\includegraphics[width=1.0 \hsize ]{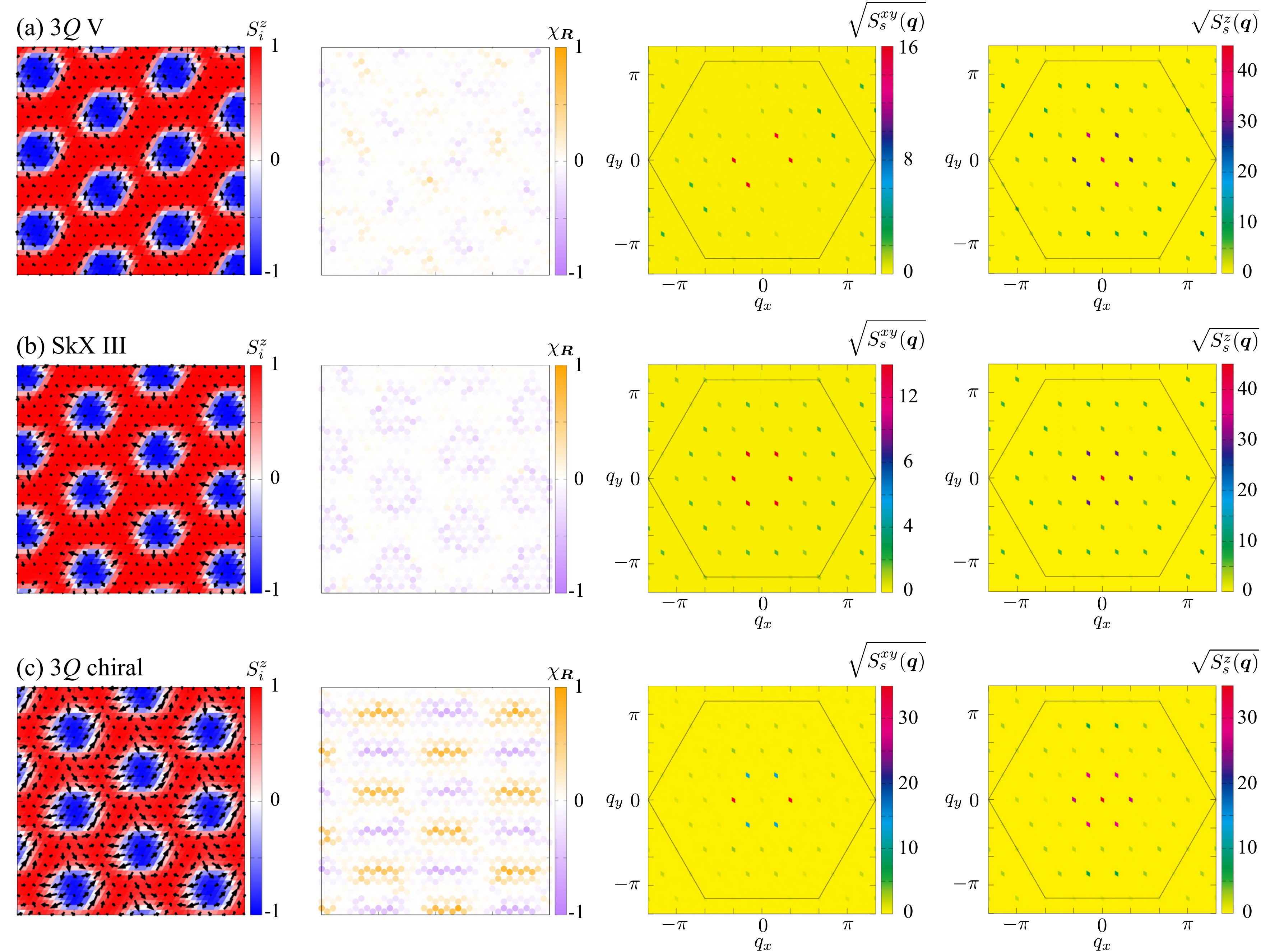} 
\caption{
\label{fig: spin_A-Hz_2}
(Left panel) Snapshots of the spin configurations in (a) the triple-$Q$ (3$Q$) V state for $A=0.5$ and $H^z=0.75$, (b) SkX III for $A=0.5$ and $H^z=0.9$, and (c) 3$Q$ chiral state for $A=0.355$ and $H^z=0.625$. 
The contours show the $z$ component of the spin moment, while the arrows represent the $xy$ components of the spin moment. 
(Middle left panel) Snapshots of the scalar chirality configurations; $\chi_{\bm{R}}=\bm{S}_{j}\cdot (\bm{S}_k \times \bm{S}_l)$. 
(Middle right and right panels) The square root of the $xy$ and $z$ components of the spin structure factor, respectively. 
The hexagons with a solid line in the right two columns show the first Brillouin zone.
}
\end{center}
\end{figure*}

\begin{figure}[t!]
\begin{center}
\includegraphics[width=0.8 \hsize ]{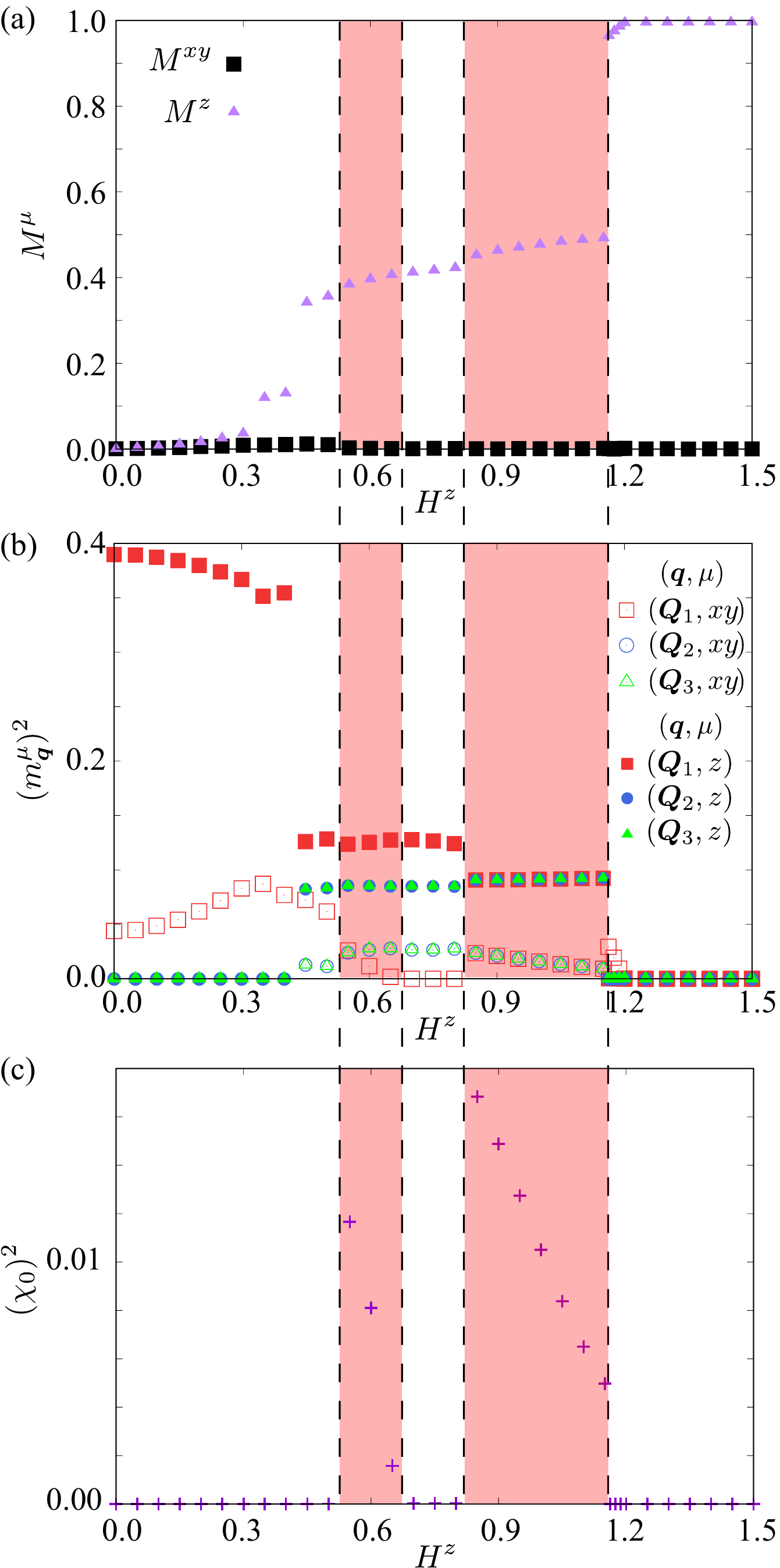} 
\caption{
\label{fig: Mq_A-Hz}
$H^z$ dependences of (a) $M^{xy}$ and $M^z$, (b) $(m^{\mu}_{\bm{q}})^2$ for $\mu=xy, z$ and $\bm{q}=\bm{Q}_1, \bm{Q}_2, \bm{Q}_3$, and (c) $(\chi_0)^2$ at $D=0.1$ and $A=0.5$. 
The vertical dashed lines represent the boundaries between the SkXs and the other magnetic phases.
}
\end{center}
\end{figure}

We discuss the situation where the magnetic field is applied in the out-of-plane direction by considering the model Hamiltonian $\mathcal{H}+\mathcal{H}_z^{\rm local}$ in Eqs.~(\ref{eq: Ham2}) and (\ref{eq: Ham_localHz}), i.e., $H^x=0$, where the model possesses the continuous rotational symmetry around the $z$ axis in spin space. 
In contrast to the case of $H^x \neq 0$ in Sec.~\ref{sec: In an in-plane magnetic field}, the SkX does not appear while varying $H^z$ in the absence of the single-ion anisotropy $A=0$, since both the out-of-plane DM interaction and the out-of-plane magnetic field favor the in-plane cycloidal spiral state. 
This is why we introduce the easy-axis single-ion anisotropy, which tends to stabilize the SkX in the out-of-plane magnetic field~\cite{leonov2015multiply, Lin_PhysRevB.93.064430, Hayami_PhysRevB.93.184413}. 

Figure~\ref{fig: PD_A-Hz} shows the magnetic phase diagram in the plane of $A$ and $H^z$ for fixed $D=0.1$, where the result at $D=0$ has been discussed in the previous studies~\cite{leonov2015multiply, Lin_PhysRevB.93.064430, Hayami_PhysRevB.93.184413}. 
Reflecting the interplay among $A$, $H^z$, and $D$, we obtain ten different magnetic phases including two types of the SkXs (SkX II and SkX III) in addition to the fully-polarized state. 
Among them, we discuss seven phases stabilized for small $A$ in this section. 
The spin and scalar chirality in real and momentum spaces are shown in Figs.~\ref{fig: spin_A-Hz} and \ref{fig: spin_A-Hz_2}. 
We also plot the $H^z$ dependences of $M^{\mu}$, $(m^{\mu}_{\bm{Q}_\nu})^2$ for $\mu=xy, z$, and $(\chi_0)^2$ at $A=0.5$ in Fig.~\ref{fig: Mq_A-Hz}. 
The spin configurations in three phases stabilized for large $A$ are shown in Appendix~\ref{sec: Spin configurations under strong easy-axis anisotropy}.

For small $A=0$, the single-$Q$ conical spiral state appears, whose spin configuration is characterized by a superposition of the in-plane cycloidal spiral and the $z$-component uniform magnetization (not shown). 
While increasing $A$, the single-$Q$ vertical spiral state and the triple-$Q$ IV state replace the single-$Q$ conical state in the low-field region, as shown in Fig.~\ref{fig: PD_A-Hz}. 
The single-$Q$ vertical spiral state is characterized by stripe spin configuration with the out-of-plane spiral modulation, as shown in the left panel of Fig.~\ref{fig: spin_A-Hz}(a). 
The magnitude of $(m^z_{\bm{Q}_1})^2$ is much larger than $(m^{xy}_{\bm{Q}_1})^2$, as shown in the right two panels of Fig.~\ref{fig: spin_A-Hz}(a) and Fig.~\ref{fig: Mq_A-Hz}(b). 
It is noted that there is a small uniform magnetization regarding the $xy$ component owing to the DM interaction, as shown in Fig.~\ref{fig: Mq_A-Hz}(a) and the middle right panel of Fig.~\ref{fig: spin_A-Hz}(a)
This state does not show a scalar chirality density wave, as shown in the middle left panel of Fig.~\ref{fig: spin_A-Hz}(a). 

In the triple-$Q$ IV state appearing upon increasing $H^z$, $(\bm{m}_{\bm{Q}_{2}})^2$ and $(\bm{m}_{\bm{Q}_{3}})^2$ become nonzero with equal intensities in Fig.~\ref{fig: Mq_A-Hz}(b). 
At first glance, the real-space spin configuration in the left panel of Fig.~\ref{fig: spin_A-Hz}(b) is similar to the SkX, while it is different from each other by closely looking at the in-plane spin alignment around the core at $S_i^z \simeq -1$; the in-plane spins are not rotated around the core, and hence, the winding number (vorticity) becomes zero. 
Accordingly, the triple-$Q$ IV state does not show a net scalar chirality similar to the single-$Q$ vertical spiral state, although this state accompanies the anisotropic triple-$Q$ chirality density waves with the dominant double-$Q$ components and subdominant single-$Q$ component, which is implicitly found in the middle left panel of Fig.~\ref{fig: spin_A-Hz}(b). 
This state has a small in-plane component of the uniform magnetization $M^{xy}$ similar to the single-$Q$ vertical spiral state. 
Meanwhile, the origin of $M^{xy}$ is attributed to the spiral plane change by the interplay among the exchange interaction, single-ion anisotropy, and magnetic field.
Indeed, a similar spin state denoted as the flop state has been discussed in the frustrated model without $D$, as shown in Fig.~1 in Ref.~\cite{leonov2015multiply}.
Although the stability region of this state is narrow for $D=0$, it is dramatically extended in the presence of $D$, as shown in Fig.~\ref{fig: PD_A-Hz}.
This is due to the fact that the out-of-plane DM interaction tends to favor the state with the in-plane component of the uniform magnetization, as found in the above single-$Q$ vertical spiral state.

When we increase $A$ and $H$, the triple-$Q$ IV state in the intermediate $H^z$ region shows multiple phase transitions to the other multiple-$Q$ states. 
We discuss such a phase sequence for fixed $A=0.5$, where the $H^z$ dependences of spin- and chirality-related quantities are shown in Fig.~\ref{fig: Mq_A-Hz}. 
At $A=0.5$, the triple-$Q$ IV state turns into an SkX II. 
Although the magnitudes of $(m^z_{\bm{Q}_{\nu}})^2$ are similar to each other, $(m^{xy}_{\bm{Q}_{\nu}})^2$ shows a slight change as shown in Fig.~\ref{fig: Mq_A-Hz}(b). 
This change results in the qualitative different spin configuration as shown in the left panel of Fig.~\ref{fig: spin_A-Hz}(c); the winding number around the core becomes nonzero, which gives a uniform component of the scalar chirality, as shown in the middle left panel of Fig.~\ref{fig: spin_A-Hz}(c). 
Although the in-plane spins around the core seem to be aligned in a threefold-symmetric way, such symmetry is lost as found in the right two panels of Fig.~\ref{fig: spin_A-Hz}(c), where the intensities of $(\bm{m}_{\bm{Q}_{\nu}})^2$ are different from each other [see also Fig.~\ref{fig: Mq_A-Hz}(b)]. 
Similar to the single-$Q$ vertical spiral state and the triple-$Q$ IV state, the SkX II exhibits the nonzero in-plane magnetization shown in the middle right panel of Fig.~\ref{fig: spin_A-Hz}(c), which indicates the spiral plane change owing to the DM interaction as found in the single-$Q$ vertical and triple-$Q$ IV states. 
In the region for the SkX II, the skyrmion number per magnetic unit cell becomes $\pm 1$~\cite{comment_SkXII}. 
Unlike the SkXs under the in-plane field, the degeneracy of the SkX and anti-SkX under the out-of-plane field is owing to the continuous rotational symmetry around the $z$ axis in spin space.
The SkX II changes into the triple-$Q$ V state or the SkX III depending on $A$. 

In the triple-$Q$ V state, the out-of-plane spin configuration is characterized by a similar triple-$Q$ superposition to the triple-$Q$ IV state, where the core forms the almost triangular lattice, as shown in the left panel of Fig.~\ref{fig: spin_A-Hz_2}(a).
The difference from the triple-$Q$ IV state appears in the in-plane spin configuration; 
the in-plane spin structure in the triple-$Q$ V state is almost characterized by the double-$Q$ superposition in Fig.~\ref{fig: Mq_A-Hz}(b), which results in almost zero in-plane uniform magnetization in Fig.~\ref{fig: Mq_A-Hz}(a). 

While further increasing $H^z$, the SkX III appears instead of the triple-$Q$ V state in Fig.~\ref{fig: PD_A-Hz}. 
As shown in the real-space spin and chirality configurations in the left two panels in Fig.~\ref{fig: spin_A-Hz_2}(b), the skyrmion core with a nonzero scalar chirality forms the triangular lattice; the skyrmion number is quantized as $\pm 1$. 
In contrast to the SkX II, the intensities of $(\bm{m}_{\bm{Q}_{\nu}})^2$ are equivalent for $\nu=1$-$3$, as shown in the right two panels of Figs.~\ref{fig: spin_A-Hz_2}(b) and Fig.~\ref{fig: Mq_A-Hz}(b). 
The SkX III turns into the single-$Q$ conical spiral state upon increasing $H^z$, and finally, turns into the fully-polarized state. 

In the narrow region sandwiched by the triple-$Q$ IV state and the SkX III, another triple-$Q$ state appears, which we refer to as a triple-$Q$ chiral state. 
Although the spin and scalar chirality configurations in the triple-$Q$ chiral state are similar to those in the triple-$Q$ IV state, this state exhibits a slight but nonzero net scalar chirality; the typical values of $(\chi_0)^2$ in this state are around $0.001$-$0.02$.
In other words, the positive and negative chirality distributions in the middle left panel of Fig.~\ref{fig: spin_A-Hz_2}(c) are not canceled out. 
This state is a topologically trivial state without the nonzero integer skyrmion number in contrast to the other SkX phases.

Let us discuss the difference between the result in Fig.~\ref{fig: PD_A-Hz} and that at $D=0$~\cite{hayami2022skyrmion}. 
The SkX III in the intermediate field in Fig.~\ref{fig: PD_A-Hz} might correspond to the SkX stabilized in the absence of $D$, since both spin textures keep threefold rotational symmetry. 
Meanwhile, other multiple-$Q$ instabilities toward the SkX II and the triple-$Q$ chiral state below the region of the SkX III occur only for $D \neq 0$. 
Thus, their appearance is owing to the interplay between the out-of-plane DM interaction and the easy-axis anisotropy. 
In addition, the single-$Q$ conical state, which does not appear for $D=0$, survives in the high-field region for $D\neq 0$. 
Thus, the identification of the low- and high-field phases as well as the SkX III in the $D_{3h}$ system in experiments is useful to understand the role of the out-of-plane DM interaction.  

Although the spin texture in the SkX III is similar to that in the SkX realized at $D=0$, there are differences due to the presence of the out-of-plane DM interaction. 
Such a difference is found in the intensities at higher-harmonic ordering vectors in the spin structure factor. 
As the higher-harmonic peak position affects the charge density waves with wave vector $\bm{q}$ in the form of $(\bm{S}_{\bm{q}_1}\cdot \bm{S}_{\bm{q}_2})\delta_{\bm{q}_1+\bm{q}_2,\bm{q}}$~\cite{Christensen_PhysRevX.8.041022,Hayami_PhysRevB.104.144404} and antisymmetric spin polarization in the band structure in the form of $(\bm{S}_{\bm{q}_1}\times  \bm{S}_{-\bm{q}_1})$~\cite{Hayami_PhysRevB.105.024413} ($\bm{q}_1, \bm{q}_2$ represent the ordering vectors or their higher harmonics and $\delta$ is the Kronecker delta), a different nature of transport property is expected when the spin texture is coupled to itinerant electrons.

\section{Summary}
\label{sec: Summary}

To summarize, we have investigated the possibility of the SkX under the point group $D_{3h}$ by focusing on the role of the out-of-plane DM interaction. 
We constructed the low-temperature magnetic phase diagrams in two situations with different magnetic-field directions by performing the simulated annealing. 
In the case of the in-plane magnetic field, we found that the out-of-plane DM interaction becomes a source of inducing the SkX in the intermediate field. 
Meanwhile, we showed that the interplay between the out-of-plane DM interaction and the easy-axis anisotropy leads to two types of the SkXs in the out-of-plane magnetic field. 
We also found the instabilities toward the other multiple-$Q$ states triggered by the out-of-plane DM interaction. 
We presented the details of the spin and scalar chirality configurations in each phase, which is useful to identify the multiple-$Q$ states in experiments.

The SkX induced by the out-of-plane DM interaction is different from that by the in-plane DM interaction in terms of the vorticity; the degeneracy of the vorticity is left (lifted) under the out-of-plane (in-plane) DM interaction.
Such a degeneracy of the vorticity might lead to the domain structure of the SkX and anti-SkX, which is called the $Z$ phase against thermal fluctuations
~\cite{Okubo_PhysRevLett.108.017206,Mitsumoto_PhysRevB.105.094427}. 
It will be left for future study to investigate the effect of finite temperatures under the out-of-plane DM interaction.

The present results will be a reference to explore the SkXs and other exotic multiple-$Q$ states in the presence of the out-of-plane DM interaction.
One of the candidate lattice structures is a breathing-kagome-lattice structure belonging to the $D_{3h}$ point group.  
Moreover, the out-of-plane DM interaction appears under other point groups, such as $C_{3h}$, $D_3$, $C_{3v}$, and $C_3$. 
Thus, in spiral magnets under these point groups, there is a chance of inducing the SkXs when applying the magnetic field. 

\appendix

\section{Spin configurations under strong easy-axis anisotropy}
\label{sec: Spin configurations under strong easy-axis anisotropy}

\begin{figure}[t!]
\begin{center}
\includegraphics[width=1.0 \hsize ]{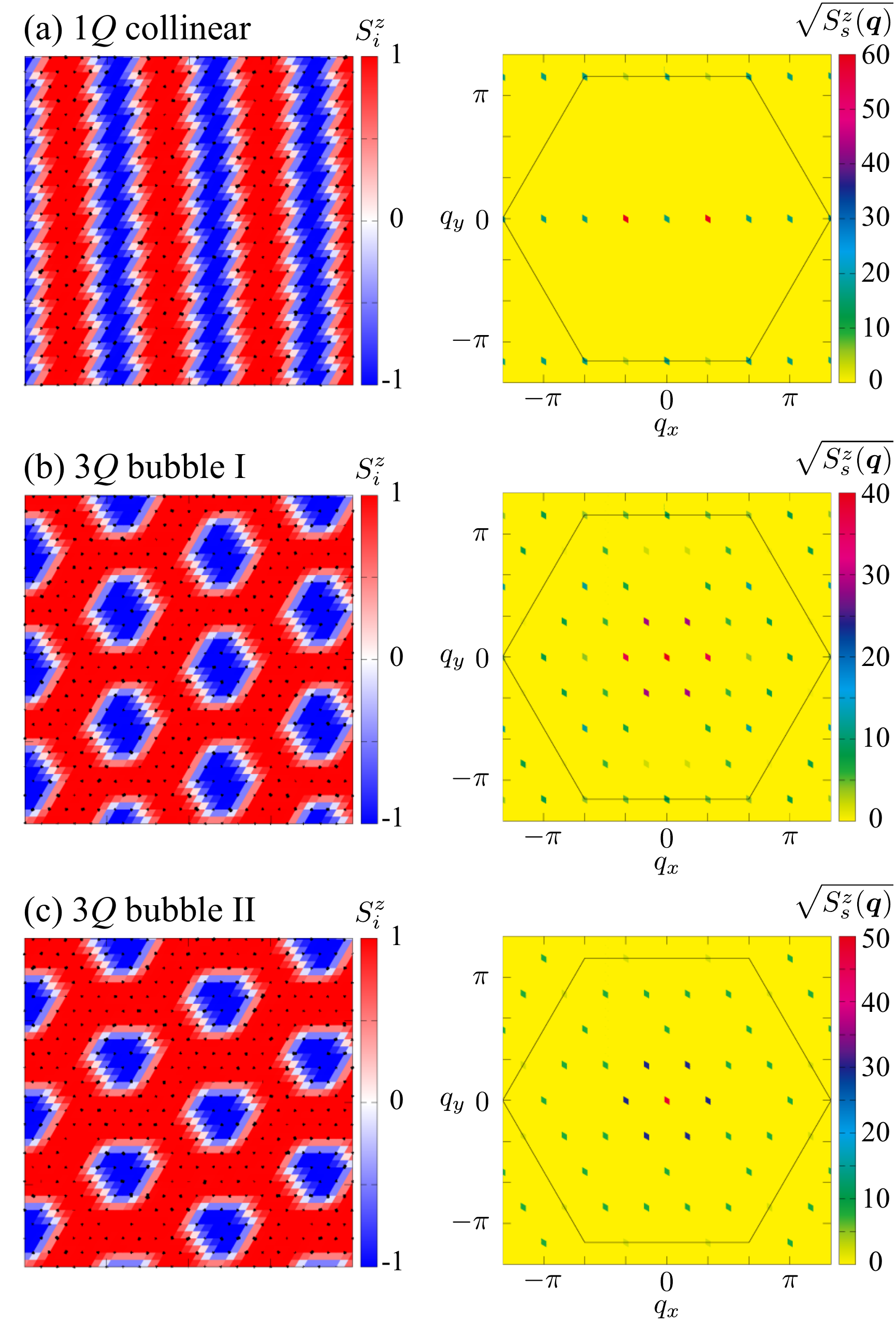} 
\caption{
\label{fig: spin_A-Hz_app}
(Left panel) Snapshots of the spin configurations in (a) the single-$Q$ (1$Q$) collinear state for $H^z=0.4$, (b) triple-$Q$ (3$Q$) bubble I for $H^z=0.6$, and (c) 3$Q$ bubble II state for $H^z=0.9$ at $A=0.595$. 
The contours show the $z$ component of the spin moment, while the arrows represent the $xy$ components of the spin moment. 
(Right panel) The square root of the $z$ component of the spin structure factor, where the hexagon with a solid line shows the first Brillouin zone.
}
\end{center}
\end{figure}

In this Appendix, we show the real-space spin configuration and the spin structure factor in the magnetic phases stabilized for large $A$ in Fig.~\ref{fig: PD_A-Hz} in Sec.~\ref{sec: In an out-of-plane magnetic field}. 
The data in Figs.~\ref{fig: spin_A-Hz_app}(a), \ref{fig: spin_A-Hz_app}(b), and \ref{fig: spin_A-Hz_app}(c) correspond to those in the single-$Q$ collinear state, the triple-$Q$ bubble I state, and the triple-$Q$ bubble II state, respectively. 
Since there is almost no $xy$-spin component in these spin textures, we only show the $z$ component of the spin structure factor.  
In other words, these spin textures are characterized by the collinear ones, which are not affected by the out-of-plane DM interaction; the strong easy-axis anisotropy is a source of these magnetic states. 
The difference between the triple-$Q$ bubble I and bubble II states is the intensities at the triple-$Q$ ordering vectors $\bm{Q}_\nu$: the former shows the anisotropic triple-$Q$ peak structure, while the latter shows the isotropic triple-$Q$ peak structure. 
The emergence of the bubble state has also been discussed in a similar triangular-lattice model~\cite{Hayami_PhysRevB.93.184413,Hayami_10.1088/1367-2630/ac3683,Utesov_PhysRevB.105.054435} and square-lattice model~\cite{Su_PhysRevResearch.2.013160,seo2021spin}.

\begin{acknowledgments}
This research was supported by JSPS KAKENHI Grants Numbers JP19K03752, JP19H01834, JP21H01037, JP22H04468, and by JST PRESTO (JPMJPR20L8).
R.Y. was supported by Forefront Physics and Mathematics Program to Drive Transformation (FoPM).
Parts of the numerical calculations were performed in the supercomputing systems in ISSP, the University of Tokyo.
\end{acknowledgments}

\bibliographystyle{apsrev}
\bibliography{ref}

\begin{thebibliography}{153}
\bibitem[{com({\natexlab{b}})}]{comment_SkXII}
\bibinfo{note}{The skyrmion number in the SkX II deviates from $\pm 1$ for
  large $A$ and $H$ owing to the discrete spin texture or thermal fluctuations.
  We identify the SkX II when the skyrmion number is larger (smaller) than
  $0.8$ ($-0.8$) at least.}
\end{thebibliography}

\end{document}